\documentclass[twocolumn,showpacs,preprintnumbers,amsmath,amssymb]{revtex4}

\usepackage{graphicx}
\usepackage{dcolumn}
\usepackage{bm}
\usepackage{verbatim}
\usepackage{accents}
\newcommand{\e}{\epsilon}
\newcommand{\vecv}{\mathbf{v}}
\newcommand{\vecq}{\mathbf{q}}

\newcommand{\vecP}{\mathbf{P}}
\newcommand{\vecI}{\mathbf{I}}
\newcommand{\vecw}{\mathbf{w}}
\newcommand{\vecC}{\mathbf{C}}
\newcommand{\vecx}{\mathbf{x}}

\begin{document}

\title{Conservation-dissipation formalism of irreversible thermodynamics}

\author{Yi Zhu}
\author{Liu Hong}
\author{Zaibao Yang}
\author{Wen-An Yong}
\email[]{wayong@tsinghua.edu.cn}
\affiliation{%
Zhou Pei-Yuan Center for Applied Mathematics, Tsinghua University, Beijing 100084, China
}%

\date{\today}

\begin{abstract}
We propose a conservation-dissipation formalism (CDF) for coarse-grained descriptions of irreversible processes. This formalism is based on a stability criterion for non-equilibrium thermodynamics. The criterion ensures that non-equilibrium states tend to equilibrium in long time. As a systematic methodology, CDF provides a feasible procedure in choosing non-equilibrium state variables and determining their evolution equations. The equations derived in CDF have a unified elegant form. They are globally hyperbolic, allow a convenient definition of weak solutions, and are amenable to existing numerics. More importantly, CDF is a genuinely nonlinear formalism and works for systems far away from equilibrium. With this formalism, we formulate novel thermodynamics theories for heat conduction in rigid bodies and non-isothermal compressible Maxwell fluid flows as two typical examples. In these examples, the non-equilibrium variables are exactly the conjugate variables of the heat fluxes or stress tensors. The new theory generalizes Cattaneo's law or Maxwell's law in a regularized and nonlinear fashion.
\end{abstract}

\pacs{05.70.Ln,51.30.+i,05.60.Cd}
\maketitle
{\em Introduction.\;}
Irreversible thermodynamics is a systematic methodology for mathematical modeling of irreversible phenomena. It has been successfully applied to many problems such as heat transfers, complex fluid flows, chemical reactions, \emph{etc.} \cite{GM,NT_2008}. As a coarse-grained theory, irreversible thermodynamics aims at determining the dynamics of non-equilibrium processes.

In general, a physical process obeys some conservation laws such as those of mass, momentum and energy. These conservation laws are expressed locally as
\begin{equation}\label{GCL}
\partial_t u+ \sum_{j=1}^3 \partial_{x_j} f_j=0.
\end{equation}
Here $u=u(t,x)\in\mathbb{R}^n$ represents conserved variables depending on the time and spatial coordinates $(t,x)$, $x=(x_1,x_2,x_3)$, and $f_j$ is the corresponding flux along the $x_j$-direction. If $f_j$ is given in terms of the conserved variables, the system \eqref{GCL} becomes closed. In this case, the system is considered to be in local equilibrium and $u$ is also referred to as equilibrium variables.
However, very often $f_j$ depends on some extra variables in addition to the conserved ones. The extra variables characterize non-equilibrium features of the system under consideration, called non-equilibrium or dissipative variables, and their choice is not unique. Thus, choosing suitable non-equilibrium variables and determining their evolution equations are the fundamental task of irreversible thermodynamics.

There have been no well-accepted rules for choosing the non-equilibrium variables and determining the evolution equations. Consequently, irreversible thermodynamics has many ``schools", such as Classical Irreversible Thermodynamics (CIT), Extended Irreversible Thermodynamics (EIT), Internal Variables Thermodynamics, Rational Thermodynamics, GENERIC (General Equation for Non-Equilibrium Reversible-Irreversible Coupling) and so on \cite{NT_2008,Many_School,EIT_book,Truesdell_book,GO,BET_book}. Each of them has its own way for choosing the non-equilibrium variables and deriving the corresponding evolution equations.

CIT, as developed by Onsager, Prigogine and many others, is a well recognized theory in the classical hydrodynamic regime \cite{GM}. It is based on the local equilibrium hypothesis. In CIT, fluxes $f_j$ are determined in terms of the conserved variables and their spatial derivatives. Typical examples are Newton's law of viscosity and Fourier's law of heat conduction. In this way, CIT leads to evolution partial differential equations (PDEs) of second-order. Although it is quite successful in modeling a wide class of real phenomena, CIT is not adequate when studying processes with long relaxation times such as heat propagation at low temperatures, polymeric fluid flows and so on \cite{NT_2008,EIT_book}.

The inadequacy of CIT has motivated various extended theories beyond the local equilibrium hypothesis. EIT is such a typical theory. It chooses the dissipative fluxes as non-equilibrium variables and pre-specifies a generalized entropy depending on both the conserved variables and dissipative fluxes \cite{EIT_book, Muller_book}. Then the evolution equations of the dissipative fluxes are derived from balance equations of the entropy. This theory leads to relaxation-type phenomenological laws such as Cattaneo's law of heat conduction and Maxwell's law of viscoelasticity. Together with the conservation laws, these phenomenological laws form a system of first-order evolution PDEs. Although EIT works for the long relaxation phenomena to a certain extent, it might not be adequate for systems far away from equilibrium \cite{BET_book, LRV_08,LY}. Moreover, the well-posedness (hyperbolicity) of the resultant governing equations does not seem clear.

On the other hand, it is well recognized that hyperbolicity is a substantial requirement for systems of first-order PDEs to be well-posed \cite{Serre}. Although there are some discussions on the hyperbolicity of the resultant PDEs in this field (e.g., \cite{Muller_book}), no extended theory till now can lead to globally hyperbolic governing equations. More seriously, the existing extended theories have not paid much attention to the corresponding short-relaxation-time limit which is closely related to the compatibility with CIT. It was shown in \cite{Yong_book} that a well-behaved relaxation-time limit requires some deliberate structural stability conditions imposing on the PDEs. For these and other reasons  \cite{Many_School,LRV_08,Garcia-Colin_95}, we consider irreversible thermodynamics to be a field in progress rather than an established edifice.

In this work, we propose a conservation-dissipation formalism (CDF) for choosing the non-equilibrium variables and determining their evolution equations. This formalism is based on the entropy dissipation condition \cite{Yong_08} ensuring that the non-equilibrium states tend to equilibrium in long time. The condition also guarantees the compatibility of the expected theory with CIT. In CDF, the non-equilibrium variables are carefully chosen so that the dissipation condition is fulfilled.
The evolution equations of the chosen non-equilibrium variables can be easily obtained from the balance equation of the entropy.

In so doing, our CDF successfully removes the blemishes of EIT mentioned above. Specifically, the resultant governing equations are automatically globally hyperbolic, the dependence of the entropy on the non-equilibrium variables is not restricted to quadratic forms, and a dissipation matrix is naturally introduced to characterize complicated nonlinear dissipation mechanisms. Moreover, our governing equations have a unified elegant form. This form is very amenable to modern mathematical theories \cite{Dafermos} on systems of first-order PDEs and to conventional numerics. For instance, weak solutions can be defined conveniently with this form. All of these advantages make us believe CDF to be promising and of great values in applications.

{\em Conservation-Dissipation Formalism.} In extended theories of irreversible thermodynamics, a non-equilibrium system is described by conserved variables and dissipative ones. The choice of the dissipative variables is generally not unique. Using different state variables may lead to different governing equations, and suitable state variables are expected to give simple governing equations which directly reveal physical insights of the processes. Usually, the conserved variables and their evolution are known as in \eqref{GCL}. Thus, our task is reduced to choose proper dissipative variables and to derive their evolution equations.

Here we present a new formalism to choose proper dissipative variables and to derive their evolution equations. Motivated by the mathematical theory on the system of first-order PDEs (see, e.g., \cite{Dafermos}), we will choose the dissipative variable $v\in\mathbb{R}^m$ so that the flux $f_j$ in \eqref{GCL} can be expressed as $ f_j =f_j(u, v)$
and $v=v(t, x)$ evolves according to balance laws of the form
\begin{equation}\label{Bal}
 \partial_tv+\sum_{j=1}^3 \partial_{x_j}g_j(u,v)=q(u,v).
\end{equation}
Here $g_j(u,v)$ is the corresponding flux and $q=q(u,v)$ is the nonzero source, vanishing at equilibrium.

Together with the conservation laws \eqref{GCL}, the evolution of a non-equilibrium state is governed by a system of first-order PDEs in the compact form
\begin{equation}\label{Cano_form}
  \partial_tU+\sum_{j=1}^3 \partial_{x_j}F_j(U)=Q(U),
\end{equation}
where
\begin{equation*}
U=\left(\begin{array}{c}u\\v\end{array}\right),\quad F_j(U)=\left(\begin{array}{c}f_j(U)\\g_j(U)\end{array}\right),\quad Q(U)=\left(\begin{array}{c}0\\q(U)\end{array}\right).
\end{equation*}
Note that not every thermodynamic variable can evolve in such a balance form, while so do the densities of extensive state variables generally \cite{BMB}. Actually, many classical systems allow such a set of state variables \cite{Yong_08}.

Notice that, in many applications, the dissipative variables evolve much faster than the conserved ones. Namely, the time scale for $v$ to reach stationary, referred to as the relaxation time, is much smaller than that for $u$. Mathematically, this means that the source term is of the form $q(u,v)=\frac{1}{\e}\tilde{q}(u,v)$ with $\e\ll1$ proportional to the relaxation time. It is natural to require that the whole system should have a well-behaved limit as $\e$ goes to zero. In this limit, the dissipative variable $v$ would be a function of the equilibrium variable $u$ or its derivatives if higher-order asymptotic expansions are included. This is exactly the regime where CIT is valid. To have a well-behaved limit, some mathematical structural conditions are required and found by Yong in \cite{Yong_book, Yong_08, Yong_ARMA_04}. One set of the sufficient conditions read as:
\begin{enumerate}
\item There is a strictly concave smooth function $\eta=\eta(U)$, called entropy, such that $\eta_{UU}\cdot F_{jU}(U)$ is symmetric for each $j$ and for all $U=(u, v)$ under consideration;
\item There is a positive definite matrix $M(U)$, called dissipation matrix, such that $q(U)=M(U)\cdot\eta_v(U).$
\end{enumerate}
Here the subscript stands for the partial derivative with respect to this subscript, for instance $\eta_v =\frac{\partial \eta}{\partial v}$ and $\eta_{UU}=\frac{\partial^2 \eta}{\partial U^2}$.

Balance equation \eqref{Cano_form} together with the two structural conditions above will be referred to as conservation-dissipation formalism (CDF). We will show with examples later how CDF guides us to choose the dissipative variable $v$ and to determine the corresponding fluxes $g_j(u,v)$ in \eqref{Bal}. Note that the source has already been proposed in the form $q(U)=M(U)\cdot\eta_v(U)$.

We conclude this part with some explanations about the two conditions above. The first one is the well-known entropy condition for hyperbolic conservation laws \cite{Dafermos, FL_71, RS}. It corresponds to the classical thermodynamics stability. This condition ensures that the expected system \eqref{Cano_form} is globally symmetrizable hyperbolic. It implies that there is a function $J_j=J_j(U)$ such that
\begin{equation}\label{4}
\eta_U\cdot F_{jU}=J_{jU}.
\end{equation}
This relation gives a restriction for the flux $g_j(u,v)$. Moreover, equation \eqref{4} leads to
\begin{align*}
\partial_t \eta&=-\sum_{j=1}^3\eta_U \cdot\partial_{x_j}F_j+\eta_v\cdot q\nonumber\\
&=-\sum_{j=1}^3\partial_{x_j}J_j+\sigma
\end{align*}
with the entropy production $\sigma=\eta_v\cdot M(U)\cdot \eta_v\ge0.$
Here the second condition has been used. Thus, the second law of thermodynamic is respected automatically by system \eqref{Cano_form} constructed via CDF.

The second condition is a nonlinearization of the celebrated Onsager reciprocal relation for scalar processes \cite{GM}. Together with the first condition, it can be regarded as a stability criterion for non-equilibrium thermodynamics. This criterion ensures that the states far away from equilibrium tend to equilibrium in long time \cite{Yong_JDE}.
Remark that the dissipation matrix may depend on the non-equilibrium variables as well as the conserved ones, while the Onsager relation only allows the dependence on the conserved variables. It was shown in \cite{Yong_08} why the dissipation matrix $M=M(U)$ must be positive instead of semi-positive definite. In fact, this positive definiteness guarantees that $\eta_v(u, v)=0$ whenever $q(u,v)=0$. This means that the local equilibrium states are those attaining the maximum of the entropy with respect to the non-equilibrium variables.

{\em Heat conduction in rigid bodies.\;}As the first example, we consider heat conduction in rigid bodies. This process obeys the conservation law of energy:
\begin{equation}\label{Energy}
\partial_t u+\nabla\cdot\vecq=0,
\end{equation}
where $u$ is the internal energy and  $\vecq$ is the corresponding heat flux. This equation is not closed since the flux $\vecq$ is unknown. Our task is to close this equation.

Unlike EIT where $\vecq$ is simply added to the state space, our CDF introduces a non-equilibrium variable $\vecw$ with the size of $\vecq$, which will be determined later. The whole state space is now given by state variables $(u, \vecw)$.

As in EIT, we specify a strictly concave function $s=s(u,\vecw)$ as the entropy for the process. In order to be compatible with the equilibrium  thermodynamics, we define the non-equilibrium temperature $\theta$ with
$$
\theta^{-1}=s_u (u, \vecw).
$$
Moreover, we refer to equation \eqref{Energy} and the generalized Gibbs relation
\begin{equation}\label{Gibbs1}
ds = s_u d u+ s_\vecw\cdot d\vecw,
\end{equation}
and deduce the evolution of the entropy:
\begin{align*}
\partial_t s=&-s_u\nabla\cdot\vecq+s_\vecw\cdot\partial_t \vecw\\=&
-\nabla\cdot(s_u\vecq)+\vecq\cdot\nabla s_u+s_\vecw\cdot\partial_t \vecw\\
=&-\nabla\cdot \mathbf{J}+\sigma.
\end{align*}
Here $\mathbf{J}=\theta^{-1}\vecq$ is the entropy flux and $$\sigma=s_\vecw\cdot\partial_t\vecw+\vecq\cdot\nabla\theta^{-1}$$
is the entropy production.

CDF suggests to choose $\vecq=s_\vecw(u, \vecw)$ and
\begin{equation}\label{Eq_w}
\partial_t\vecw+\nabla\theta^{-1}=\mathbf{M}\cdot\vecq,
\end{equation}
where the dissipation matrix $\mathbf{M}=\mathbf{M}(u,\vecw)$ is positive definite.
Equations \eqref{Energy} and \eqref{Eq_w} together compose a system of first order PDEs in the form \eqref{Cano_form} with
\begin{equation*}
U= \left(\begin{array}{c}
   u\\
  \vecw
  \end{array}\right), \,
  \sum_j \partial_{x_j}F_j(U)=\nabla\cdot\left(\begin{array}{c}
 \vecq\\
 \theta^{-1}\vecI
      \end{array}\right),\,Q(U)= \left(\begin{array}{c}
    0\\
    \mathbf{M}\cdot
    \vecq
   \end{array}\right)
\end{equation*}
where $\vecI$ is the 3$\times$3 identity matrix.

From the above procedure, we see that the non-equilibrium variable $\vecw$ is conjugated to the heat flux $\vecq$ with respect to the pre-specified entropy. Thanks to the strict concavity of $\eta=\eta(u, \vecw)$, the non-equilirium variable $\vecw$ can be globally expressed in terms of $\vecq$ and $u$ \cite{FL_71}. Compared to EIT where directly derived was the evolution equation of the flux $\vecq$, CDF gives equation \eqref{Eq_w} which ensures that the final system is symmetrizable hyperbolic. Moreover, there are many freedoms in choosing the entropy $s$ and the dissipation matrix $M$. Indeed, CDF does not impose any further restrictions
except that the entropy is strictly concave and the dissipation matrix is positive-definite. The exact expressions of $s$ and $\mathbf{M}$ depend on the specific problem to be studied.

A simple choice of the two quantities above is
\begin{equation}
s(u,\vecw)=s_0(u) - \frac{1}{2\alpha_0}|\vecw|^2, \quad \mathbf{M}=\frac{1}{\lambda\theta^2}\vecI,
\end{equation}
where $s_0(u)$ is the equilibrium entropy, $\alpha_0$ is a constant related to the relaxation time, and $\lambda$ is the heat conduction coefficient. With this choice, we have $\vecq = -\vecw/\alpha_0$ and equation \eqref{Eq_w} reduces to Cattaneo's law
\begin{equation*}
\alpha_0\partial_t \vecq-\nabla\theta^{-1}=-\frac{1}{\lambda\theta^2}\vecq
\end{equation*}
and whose stationary limit ($\alpha_0\rightarrow 0$) gives Fourier's law $\vecq = -\lambda\nabla\theta$.
Thus, equation \eqref{Eq_w} can be regarded as a nonlinear generalization of Cattaneo's law. See \cite{LRV_08} for similar nonlinear extensions. Moreover, the stationary limit of equation \eqref{Eq_w} with general entropy functions reads as
\[
\vecq=\mathbf{M}^{-1}\cdot\nabla\theta^{-1}.
\]
This is a generalization of Fourier's law and may describe non-isotropic and nonlinear heat conduction.

{\em One-component fluids.\;} Consider a compressible one-component fluid. Without external forces, the conservation laws of mass, momentum and energy for such electrically neutral fluids read as
\begin{subequations}\label{CLE}
\begin{eqnarray}
&&\partial_t\rho + \nabla\cdot(\rho \vecv)  = 0,\label{rho} \\
&&\partial_t(\rho \vecv) + \nabla\cdot(\rho \vecv\otimes \vecv + \vecP)=0,\\
&&\partial_t(\rho e) + \nabla\cdot(\vecv\rho e+\vecq+\vecP\cdot\vecv)  = 0.
\end{eqnarray}
\end{subequations}
Here $\rho$ is the fluid density, $\vecv$ is the velocity, $e$ is the specific total energy, and the symbol $\otimes$ represents the tensor product. Since the stress tensor $\vecP$ and the heat flux $\vecq$ are unspecified, the above equations are needed to be closed.

To close these equations, we introduce two non-equilibrium state variables $\vecw$ and $\vecC$ which have the respective sizes of the vector $\vecq$ and tensor $\vecP$. Recall that such a system in equilibrium usually has a specific entropy $s_0=s_0(\nu, u)$
with $\nu = 1/\rho$ the specific volume and $u=e-|\vecv|^2/2$ the specific internal energy. Now we assume that the non-equilibrium system under consideration possesses a generalized specific entropy
$$
s = s(\nu, u, \vecw, \vecC)
$$
depending on the non-equilibrium variables $(\vecw, \vecC)$ as well as the classical ones $(\nu, u)$.
This specific entropy corresponds to the entropy density $\eta$ in CDF with the relation
\[
\eta=\eta(\rho,\rho\vecv,\rho e,\rho \vecw,\rho\vecC)=\rho s(1/\rho, e-|\vecv|^2/2, \vecw, \vecC).
\]
It is not difficult to show that the concavity of $\eta$ in its arguments is equivalent to that  of $s$ in its own arguments. In what follows, we will use $s$ instead of $\eta$, in order to easily compare with the classical calculations \cite{GM}.

Accordingly, we define the non-equilibrium temperature $\theta$ and the non-equilibrium
thermodynamic pressure $\pi$ as
\[
\theta^{-1} :=s_u(\nu, u, \vecw, \vecC), \quad \theta^{-1}\pi :=s_\nu(\nu, u, \vecw, \vecC).
\]
For convenience, we exempt the thermodynamic pressure $\pi$ from the stress $\vecP$. Namely, set $\boldsymbol\tau=\vecP-\pi\vecI$ which accounts for possible dissipative effects such as viscosity.
Moreover, we have the generalized Gibbs relation
\begin{equation}\label{Gene_Gibbs}
\hbox{d}s=\theta^{-1}\left[\pi\hbox{d}\nu +d u\right]+s_\vecw\cdot\hbox{d}\vecw+s_\vecC^T:\hbox{d}\vecC.
\end{equation}
where the superscript $T$ denotes the transpose and the colon $:$ stands for the double contraction of two second-order tensors:  $\mathbf{A}:\mathbf{B}=\sum_{i,j}A_{ij}B_{ji}$.

Next, we introduce a differential operator $\mathcal{D}$ acting on a function $f=f(\vecx,t)$ as $\mathcal{D}f:=(\rho f)_t+\nabla\cdot(\vecv\rho f).$
Thanks to the continuity equation \eqref{rho}, it is easy to see that $\mathcal{D}f=\rho (f_t+\vecv\cdot\nabla f)$ and thereby is Galilean invariant. From the equations \eqref{CLE}--\eqref{Gene_Gibbs}
we calculate the balance equation for the specific entropy as follows
\begin{align*}
&\eta_t+\nabla\cdot(\vecv\eta)\equiv\mathcal{D}s\\=
&\theta^{-1}\left[\pi\nabla\cdot\vecv-\nabla\cdot\vecq-\vecP^T:\nabla\vecv\right] \\
&+s_\vecw\cdot\mathcal{D} \vecw+s_{\vecC}^T:\mathcal{D} \vecC\\
=&-\nabla\cdot(\theta^{-1}\vecq)
+(s_\vecw\cdot\mathcal{D}\vecw+\vecq\cdot\nabla \theta^{-1})\\
&+(s_\vecC^T:\mathcal{D} \vecC-\theta^{-1}\boldsymbol\tau^T:\nabla\vecv)\\
=&-\nabla\cdot\mathbf{J}+\sigma
\end{align*}
Here $\mathbf{J}=\theta^{-1}\vecq$ is the entropy flux and
\begin{equation}\label{Entrop_Pro}
\sigma=(s_\vecw\cdot\mathcal{D}\vecw+\vecq\cdot\nabla \theta^{-1})+(s_\vecC^T:\mathcal{D} \vecC-\theta^{-1}\boldsymbol\tau^T:\nabla\vecv)
\end{equation}
is the entropy production.

Having the expression of the entropy production, we refer to CDF and choose $\vecq= s_\vecw, ~\boldsymbol\tau = \theta s_\vecC$,
and
\begin{equation}\label{13}
  \left(\begin{array}{c}
\partial_t(\rho\vecw)+\nabla\cdot(\rho\vecv\otimes\vecw)+\nabla\theta^{-1}\\
\partial_t(\rho\vecC)+\nabla\cdot(\rho\vecv\otimes\vecC)-\nabla\vecv
  \end{array}\right)=\mathbf{M}\cdot  \left(\begin{array}{c}
    \vecq\\
    \theta^{-1}\boldsymbol\tau
  \end{array}\right)
\end{equation}
with $\mathbf{M}=\mathbf{M}(\rho, u, \vecw, \vecC)$ positive definite.
Consequently, the final closed system of governing equations is of the balance form \eqref{Cano_form}. 


Up to now, we have not assumed the symmetry of the stress tensor.
If the stress tensor is symmetric, we will take the non-equilibrium tensor $\vecC$ to be symmetric. In this case, the previous calculations are still valid but lead to
\begin{equation}\label{14}
  \left(\begin{array}{c}
\partial_t(\rho\vecw)+\nabla\cdot(\rho\vecv\otimes\vecw)+\nabla\theta^{-1}\\
\partial_t(\rho\vecC)+\nabla\cdot(\rho\vecv\otimes\vecC)-\frac{1}{2}(\nabla\vecv+\nabla\vecv^T)
  \end{array}\right)=\mathbf{M}\cdot  \left(\begin{array}{c}
    \vecq\\
    \theta^{-1}\boldsymbol\tau
  \end{array}\right),
\end{equation}
instead of equation \eqref{13}.

Furthermore, we give some choices of the specific entropy function and the dissipation matrix. For convenience, we define
\begin{eqnarray*}
\accentset{\bullet}{\mathbf{A}}=\frac{1}{3}\hbox{Tr}(\mathbf{A})\mathbf{I}, \quad \accentset{\circ}{\mathbf{A}}=\frac{1}{2}(\mathbf{A}+\mathbf{A}^T)-\frac{1}{3}\hbox{Tr}(\mathbf{A})\mathbf{I}
\end{eqnarray*}
for square matrix $\mathbf{A}$, where $\hbox{Tr}(\mathbf{A})$ is the trace of $\mathbf{A}$. Note that $\accentset{\bullet}{\mathbf{A}}$ and $\accentset{\circ}{\mathbf{A}}$ are orthogonal in the sense of double contraction.

For the generalized entropy, we take
$$
s=s_0(\nu, u) - \frac{1}{2\nu\alpha_0}|\vecw|^2 - \frac{1}{2\nu\alpha_1}|\accentset{\bullet}\vecC|^2
 - \frac{1}{2\nu\alpha_2}|\accentset{\circ}\vecC|^2,
$$
where $s_0(\nu, u)$ is the strictly concave equilibrium entropy and $\alpha_0, \alpha_1$ and $\alpha_2$ are three positive parameters related to the relaxation times,
and the dissipation matrix is chosen so that
$$
\mathbf{M}\cdot  \left(\begin{array}{c}
    \vecq\\[3mm]
    \theta^{-1}\boldsymbol\tau
  \end{array}\right)= \left(\begin{array}{c}
\frac{1}{\theta^2\lambda}\vecq\\[3mm]
\frac{\accentset{\bullet}{\boldsymbol\tau}}{\xi}
+\frac{\accentset{\circ}{\boldsymbol\tau}}{\kappa}
\end{array}\right)
  $$
with $\xi$ and $\kappa$  positive viscosity parameters. Obviously, the specific entropy $s$ thus chosen is strictly concave and the dissipation matrix $\mathbf{M}$ is positive-definite.

Now we have
$$
\vecq = - \frac{\rho\vecw}{\alpha_0}, \quad \theta^{-1}\boldsymbol\tau = -\frac{\rho\accentset{\bullet}\vecC}{\alpha_1} - \frac{\rho\accentset{\circ}\vecC}{\alpha_2}.
$$
The evolution equations \eqref{14} become
\begin{eqnarray*}
&&\alpha_0\big[\partial_t \vecq +\nabla\cdot(\vecv\otimes\vecq)\big] -\nabla\theta^{-1}=-\frac{\vecq}{\theta^2\lambda},\\
&&\alpha_1\big[\partial_t(\theta^{-1}\accentset{\bullet}{\boldsymbol\tau})+
\nabla\cdot(\theta^{-1}\vecv\otimes\accentset{\bullet}{\boldsymbol\tau})\big]+
\accentset{\bullet}{\nabla\vecv}=-\frac{\accentset{\bullet}{\boldsymbol\tau}}{\kappa},\\
&&\alpha_2\big[\partial_t(\theta^{-1}\accentset{\circ}{\boldsymbol\tau})+
\nabla\cdot(\theta^{-1}\vecv\otimes\accentset{\circ}{\boldsymbol\tau})\big]+
\accentset{\circ}{\nabla\vecv}=-\frac{\accentset{\circ}{\boldsymbol\tau}}{\xi}.
\end{eqnarray*}
Here we have used the aforesaid orthogonal decomposition for the equation of the symmetric tensor $\vecC$ in \eqref{14}. These equations correspond to Cattaneo's law and Maxwell's laws of viscoelasticity, respectively. They may give a reasonable description of non-isothermal compressible Maxwell fluid flows.

In the limit as the $\alpha$'s go to zero, we arrive at
\[\vecq=-\lambda\nabla \theta, \quad \boldsymbol\tau=-\xi\accentset{\circ}{\nabla\vecv}-\kappa\accentset{\bullet}{\nabla\vecv}.
\]
These are the classical Fourier-Newton-Stokes' constitutive relations, provided that $\lambda, \xi$ and $\kappa$ are all independent of the non-equilibrium variables.
When $\lambda, \xi$ or $\kappa$ depend on the non-equilibrium variables, the above relations may describe generalized Newtonian fluids \cite{BAH}. To see this, we assume $\kappa=\xi=\mu_0|\boldsymbol\tau|^\alpha$ for simplicity and set $\dot{\boldsymbol\gamma}=\accentset{\circ}{\nabla\vecv}+ \accentset{\bullet}{\nabla\vecv}=\frac{1}{2}(\nabla\vecv+\nabla\vecv^T)$. Thus, we deduce from the second relation above that
$$
|\boldsymbol\tau| = \mu_0|\boldsymbol\tau|^\alpha|\dot{\boldsymbol\gamma}|
$$
and thereby $|\boldsymbol\tau|=(\mu_0|\dot{\boldsymbol\gamma}|)^\frac{1}{1-\alpha}$ if $\alpha\ne1$. Set $n = \frac{1}{1-\alpha}$. Then we obtain the constitutive equation
$$
\boldsymbol\tau = -\mu_0^n|\dot{\boldsymbol\gamma}|^{n-1}\dot{\boldsymbol\gamma}.
$$
for power-law fluids with index $n$ and consistency coefficient $\mu_0^n$.

{\em Summary and discussions.\;}In this work, we propose a conservation-dissipation formalism (CDF) for coarse-grained descriptions of irreversible processes. This formalism is based on a stability criterion for non-equilibrium thermodynamics. The stability means that the states far away from equilibrium tend to equilibrium in long time. It implies the compatibility of the expected theory with CIT.

Like EIT, our CDF is easily feasible. It starts with the known conservation laws of form \eqref{GCL}. The key step is to choose the non-equilibrium variables, a strictly concave (entropy) function of the non-equilibrium as well as conserved variables, and a positive-definite matrix characterizing the dissipation. The concavity corresponds to the classical thermodynamic stability and therefore the non-equilibrium variables should be of the nature of extensive variables. Once these quantities are chosen, the evolution equation for the non-equilibrium variables can be easily obtained by deriving the dynamics of the entropy.

To illustrate the above procedure, we formulate a novel thermodynamics theory for
general one-component fluid flows. In this example, the non-equilibrium variables are exactly the conjugate variables of the dissipative fluxes in EIT. The new theory is a nonlinear extension of Cattaneo's law or Maxwell's law.

Our CDF possesses some important advantages of GENERIC \cite{BET_book,GO} --- a popular formalism for irreversible thermodynamics. For instance, CDF is a genuinely nonlinear formalism, it works for systems far away from equilibrium, and the equations derived in CDF have a unified elegant form and respects the second law of thermodynamics.
However, GENERIC does not seem to provide hints to choose the non-equilibrium variables, it involves complicated bracket algebras and tedious calculations, and its mathematical foundation needs to be justified. On the other hand, our CDF is easily feasible and understandable, the resultant governing equations are globally hyperbolic and thereby well-posed for initial-value problems, and these equations allow a convenient definition of weak solutions. As a consequence of these advantages, the equations derived in CDF are very amenable to existing numerics and thus CDF is anticipated to have great values in applications.

Like other approaches including EIT and GENERIC, our CDF has many freedoms on choosing the entropy function and the dissipation matrix. These freedoms leave more flexibility to our CDF in modeling various systems in different regimes. Generally speaking, these freedoms are problem-dependent and could be reduced by compatibility considerations. For example, the dissipation matrix for the one-component fluids is required by the Galilean invariance to be independent of the fluid velocity, it is often chosen to be block-diagonal for the decoupling of tensors with different orders in view of Curie's principle, and so on. Further discussions on the reduction are beyond the scope of this paper and will be addressed in the future.
\vskip 4mm
This work was partially supported by NSFC under grant no. 11204155, Tsinghua University Initiative Scientific Research Program under grant nos. 20121087902, 20131089184.


\begin{thebibliography}{10}

\bibitem{GM}
S.~R. de~Groot and P.~Mazur.
\newblock {\em Non-Equilibrium Thermodynamics}.
\newblock North-Holland Publishing Company, Amsterdam, 1962.

\bibitem{NT_2008}
G.~Lebon, D.~Jou, and J.~Casas-V¡äazquez.
\newblock {\em Understanding Non-equilibrium Thermodynamics: Foundations,
  Applications, Frontiers}.
\newblock Springer-Verlag, Berlin, 2008.

\bibitem{Many_School}
W.~Muschik.
\newblock Why so many ``schools" of thermodynamics?
\newblock {\em Forsch Ingenicurwes}, 71:149--161, 2007.

\bibitem{EIT_book}
D.~Jou, J.~Casas-V\'{a}zquez, and G.~Lebon.
\newblock {\em Extended Irreversible Thermodynamics}.
\newblock Springer, New York, 2010.

\bibitem{Truesdell_book}
C.~Truesdell.
\newblock {\em Rational Thermodynamics}.
\newblock Springer-Verlag, New York, 1984.

\bibitem{GO}
M.~Grmela and H.~C. \"Ottinger.
\newblock Dynamics and thermodynamics of complex fluids. {I}. development of a
  general formalism.
\newblock {\em Phys. Rev. E}, 56(6):6620, 1997.

\bibitem{BET_book}
H.~C. \"Ottinger.
\newblock {\em Beyond Equilibrium Thermodynamics}.
\newblock Wiley-Interscience, Hoboken, 2005.

\bibitem{Muller_book}
I.~M\"uller and T.~Ruggeri.
\newblock {\em Rational Extended Thermodynamics}.
\newblock Springer Verlag, New York, 1998.

\bibitem{LRV_08}
G.~Lebon, M.~Ruggieri, and A.~Valenti.
\newblock Extended thermodynamics revisited: renormalized flux variables and
  second sound in rigid solids.
\newblock {\em J. Phys: Condens. Matter}, 20:025223, 2008.

\bibitem{LY}
E.~H. Lieb and J.~Yngvason.
\newblock The entropy concept for non-equilibrium states.
\newblock {\em Proc. Roy. Soc. A}, 469:0408, 2013.

\bibitem{Serre}
D.~Serre.
\newblock {\em Systems of Conservation Laws 1: Hyperbolicity, Entropies, Shock
  Waves}.
\newblock Cambridge University Press, London, 1999.

\bibitem{Yong_book}
W.-A. Yong.
\newblock {\em Advances in the Theory of Shock Waves}, volume~47 of {\em Prog.
  Nonlinear Differ. Eqns. Appl.}, chapter Basic aspects of hyperbolic
  relaxation systems, pages 259--305.
\newblock MA: Birkh\"auser, Boston, 2001.

\bibitem{Garcia-Colin_95}
L.S. Garc\'{\i}a-Col\'{\i}n.
\newblock Extended irreversible thermodynamics: an unfinished task.
\newblock {\em Molecular Physics}, 86:697--706, 1995.

\bibitem{Yong_08}
W.-A. Yong.
\newblock An interesting class of partial differetial equations.
\newblock {\em J. Math. Phys.}, 49:033503, 2008.

\bibitem{Dafermos}
C.~M. Dafermos.
\newblock {\em Hyperbolic Conservation Laws in Continuum Physics}.
\newblock Springer, Berlin, 2000.

\bibitem{BMB}
M.~L. Bellac, F.~Mortessagne, and G.~G. Batrouni.
\newblock {\em Equilibrium and Non-equilibrium Statistical Thermodynamics}.
\newblock Cambridge University Press, New York, 2004.

\bibitem{Yong_ARMA_04}
W.-A. Yong.
\newblock Entropy and global existence for hyperbolic balance laws.
\newblock {\em Arch. Rat. Mech. Anal.}, 172:247--266, 2004.

\bibitem{FL_71}
K.~O. Friedrichs and P.~D. Lax.
\newblock Systems of conservation equations with a convex extension.
\newblock {\em Proc. Nat. Acad. Sci.}, 68:1686--1688, 1971.

\bibitem{RS}
T.~Ruggeri and A.~Strumia.
\newblock Main field and convex covariant density for quasi-linear hyperbolic
  systems: relativistic fluid dynamics.
\newblock {\em Ann. Inst. Henri Poincare, Sect. A}, 34:65, 1981.

\bibitem{Yong_JDE}
W.-A. Yong.
\newblock Singular perturbations of first-order hyperbolic systems with stiff
  source terms.
\newblock {\em J. Diff. Equations}, 155:89--132, 1999.

\bibitem{BAH}
R.~B. Bird, R.~C. Amstrong, and O.~Hassager.
\newblock {\em Dynamics of polymeric liquids, Vol. 1: fluid mechanics}.
\newblock John Wiley \& Sons Inc., New York, 1987.

\end{thebibliography}

\end{document}